# Beam test results of a 16 ps timing system based on ultra-fast silicon detectors

N. Cartiglia[1], A. Staiano, V. Sola
*INFN, Torino, Italia*

R. Arcidiacono
*Università del Piemonte Orientale, Italia and INFN, Torino, Italia*

R. Cirio, F. Cenna, M. Ferrero, V. Monaco, R. Mulargia, M. Obertino, F. Ravera, R. Sacchi
*Università di Torino, Torino, Italia and INFN, Torino, Italia*

A. Bellora, S. Durando
*Università di Torino, Torino, Italia*

M. Mandurrino
*Politecnico di Torino, Torino, Italia*

N. Minafra
*University of Kansas, KS, USA*

V. Fadeyev, P. Freeman, Z. Galloway, E. Gkougkousis, H. Grabas, B. Gruey, C.A. Labitan, R. Losakul, Z. Luce, F. McKinney-Martinez, H. F.-W. Sadrozinski, A. Seiden, E. Spencer, M. Wilder, N. Woods, A. Zatserklyaniy
*SCIPP, Univ. of California Santa Cruz, CA 95064, USA*

G.Pellegrini, S. Hidalgo, M. Carulla, D. Flores, A. Merlos, D. Quirion
*Centro Nacional de Microelectrónica (IMB-CNM-CSIC), Barcelona, Spain*

V. Cindro, G.Kramberger, I. Mandić, M. Mikuž, M. Zavrtanik
*Jožef Stefan institute and Department of Physics, University of Ljubljana, Ljubljana, Slovenia*

*Abstract*– In this paper we report on the timing resolution obtained in a beam test with pions of 180 GeV/c momentum at CERN for the first production of 45 μm thick Ultra-Fast Silicon Detectors (UFSD). UFSD are based on the Low-Gain Avalanche Detector (LGAD) design, employing n-on-p silicon sensors with internal charge multiplication due to the presence of a thin, low-resistivity diffusion layer below the junction. The UFSD used in this test had a pad area of 1.7 $mm^2$. The gain was measured to vary between 5 and 70 depending on the sensor bias voltage. The experimental setup included three UFSD and a fast trigger consisting of a quartz bar readout by a SiPM. The timing resolution was determined by doing Gaussian fits to the time-of-flight of the particles between one or more UFSD and the trigger counter. For a single UFSD the resolution was measured to be 34 ps for a bias voltage of 200 V, and 27 ps for a bias voltage of 230 V. For the combination of 3 UFSD the timing resolution was 20 ps for a bias voltage of 200 V, and 16 ps for a bias voltage of 230 V.

[1] Corresponding author: cartiglia@to.infn.it

## 1 INTRODUCTION

We are developing an ultra-fast silicon detector (UFSD) that would establish a new paradigm for space-time particle tracking [1]. The UFSD is a single device that ultimately will measure with high precision concurrently the space (~10 µm) and time (~10 ps) coordinates of a particle.

First applications of UFSD are envisioned in upgrades of experiments at the High-Luminosity Large Hadron Collider (HL-LHC [2]), in cases where the excellent time resolution coupled with good spatial resolution helps to reduce drastically pile-up effects due to the large number of individual interaction vertices, as demonstrated in [3]. ATLAS is proposing UFSD as one of the technical options for the High Granularity Timing Detector (HGTD) located in front of the forward calorimeter (FCAL) [3]. CMS and TOTEM are considering UFSD to be the timing detectors for the high momentum - high rapidity Precision Proton Spectrometer (CT-PPS) [4]. In both cases, the UFSD would be of moderate segmentation (a few $mm^2$) with challenging radiation requirements (several $10^{15}$ $neq/cm^2$).

UFSD are thin pixelated n-on-p silicon sensors - based on the LGAD design [5][6] developed by the Centro Nacional de Microelectrónica (CNM) Barcelona - whose geometry has been optimized for precision time measurements [7]. The sensor exhibits moderate internal gain (~5-70) due to a highly doped $p^+$ region just below the n-type implants. Up to now, UFSD with a large gain range were only available with a thickness of 300 µm and their performance has been established in several beam tests and with laser laboratory measurements [1][8]. When these measurements were extrapolated to thin sensors using the simulation program Weightfield2 (WF2) [9], a timing resolution of 30 ps was predicted for a 50 µm thick sensor with gain 10 and a capacitance of 2 pF [1].

In this paper, we report on the first results of a beam test with thin UFSD. In Section 2 we will briefly describe the characteristics of the 45 µm UFSD manufactured by CNM under the auspices of the RD50 collaboration [10], including electrical characterization and charge collection study in a laboratory $^{90}$Sr β-source. We will then describe the development of a low-noise 2 GHz amplifier in Section 3, and that of a fast Cerenkov trigger counter in Section 4. Set-up and running conditions of the CERN H8 test beam area will be described in Section 5, followed by the results on the timing resolution and their interpretation in Section 6.

## 2 PROPERTIES OF THIN UFSD

The UFSD were manufactured by CNM within a "RD50 Common Project" [10]. Details of the design can be found in Ref. [6]. They are produced on 4" Silicon-on-Insulator wafers with a 45 µm thick high resistivity float zone (FZ) active layer on top of a 1 µm buried oxide and a 300 µm support wafer. The back-side contact is done through wet-etched deep access holes through the insulator. The wafers contain a variety of pad structures, with the pad area $A$ varying from 1.7 $mm^2$ to 10.2 $mm^2$. This test uses solely the 1.7 $mm^2$ pads. Three sets of wafers were produced, identical in the mask design but with a different $p^+$ dose of the gain layer to optimize the charge multiplication. The results presented in this article are obtained using the intermediate gain implant dose [6]. Capacitance-voltage (C-V) and current-voltage (I-V) measurements were performed. The C-V measurements were done with grounded guard ring and resulted in a depletion voltage below 50 V. The detector capacitance was measured to be $C_{UFSD}$ = 3.9 pF, from which the active thickness of the UFSD $w$ was derived to be $w$ = 45 µm using the relationship $w = \varepsilon_r \varepsilon_o * A/C$, where $\varepsilon_o$ is the vacuum absolute permittivity and $\varepsilon_r$ is the silicon relative permittivity. The I-V measurements of the UFSD (Figure 1) revealed a bulk leakage current of below 1 nA at 240 V with a constant guard ring current of about 1 nA and a breakdown voltage of about $V_{BD}$ = 250 V.
The gain was measured in the laboratory using the read-out board described in Section 3 and β-particles from a $^{90}Sr$ source. The gain is computed by comparing the collected charge to the charge expected from a MIP in an equivalent sensor without gain, estimated to be 0.46 fC using Ref. [11]. The gain values varied from M = 5 to 70 for bias voltages between 70 V and 250 V, respectively (Figure 1) with an estimated systematic error of 20%. This value was computed by determining the gain with several signal sources (β-particles, laser, minimum ionizing particles) and electronics (integrating and fast read-out), and using the difference between the highest and lowest value.

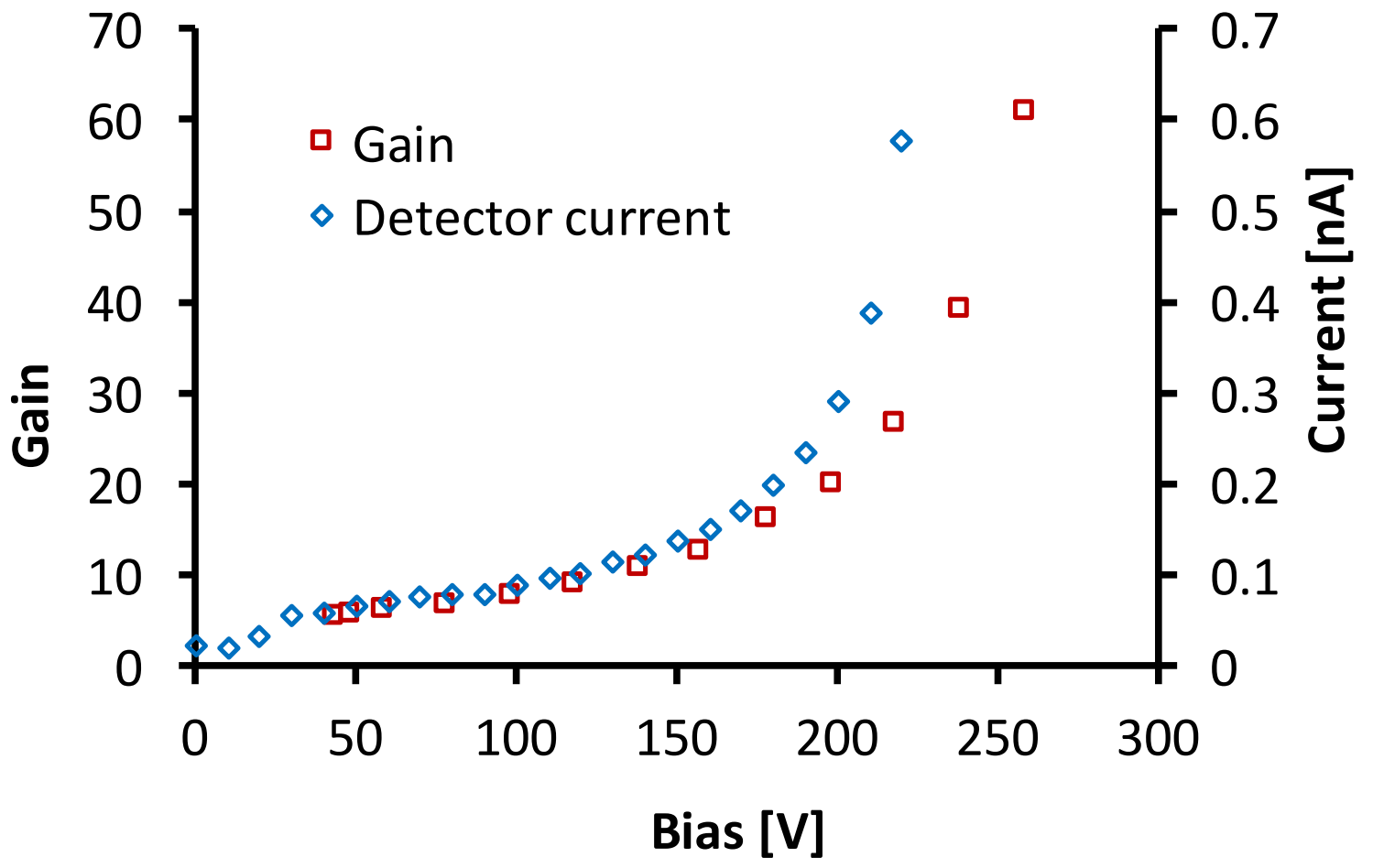

Figure 1 Bias dependence of the leakage current and the signal gain for a 45 µm-thick UFSD sensor with an area of 1.7 $mm^2$. The gain determination is subject to an overall scale uncertainty of 20%. The similar exponential behaviour of gain and sensor current below 200 V bias is due to the common charge multiplication mechanism.

As shown in Figure 1, the leakage current and the gain exhibit the same bias voltage dependence, as expected from the common charge multiplication mechanism. Above a bias of 200 V, the current shows a faster increase with the bias voltage than the gain, which might be due to the start of breakdown in the periphery of the pad, instead of the central bulk where the signal is generated.

## 3 READ-OUT BOARD

For charge collection studies in both the β-source and the beam test, the UFSD are mounted on a ~10 cm square read-out board developed at the University of California Santa Cruz (UCSC). The board uses discrete components and contains several features which allow maintaining a wide bandwidth (~ 2 GHz) and a low noise even in noisy environments: (i) by-pass capacitors located right next to the sensor, (ii) large ground planes, (iii) low impedance connections among layers, (iv) very short parallel wire-bonds to limit the inductance, and (v) self-shielding packaging using lids which snap onto the boards on both sides. The inverting amplifier uses a high-speed SiGe transistor and it has a trans-impedance of about 470 Ω. This amplifier is followed by an external commercial amplifier of gain 100 [12]. Figure 2 shows several aspects of the board: (left) the sensor with the wire bonds and the filtering capacitors, (centre) the position of the sensor with respect of the amplifier and the large ground plane around it and (right) the shielding lid.

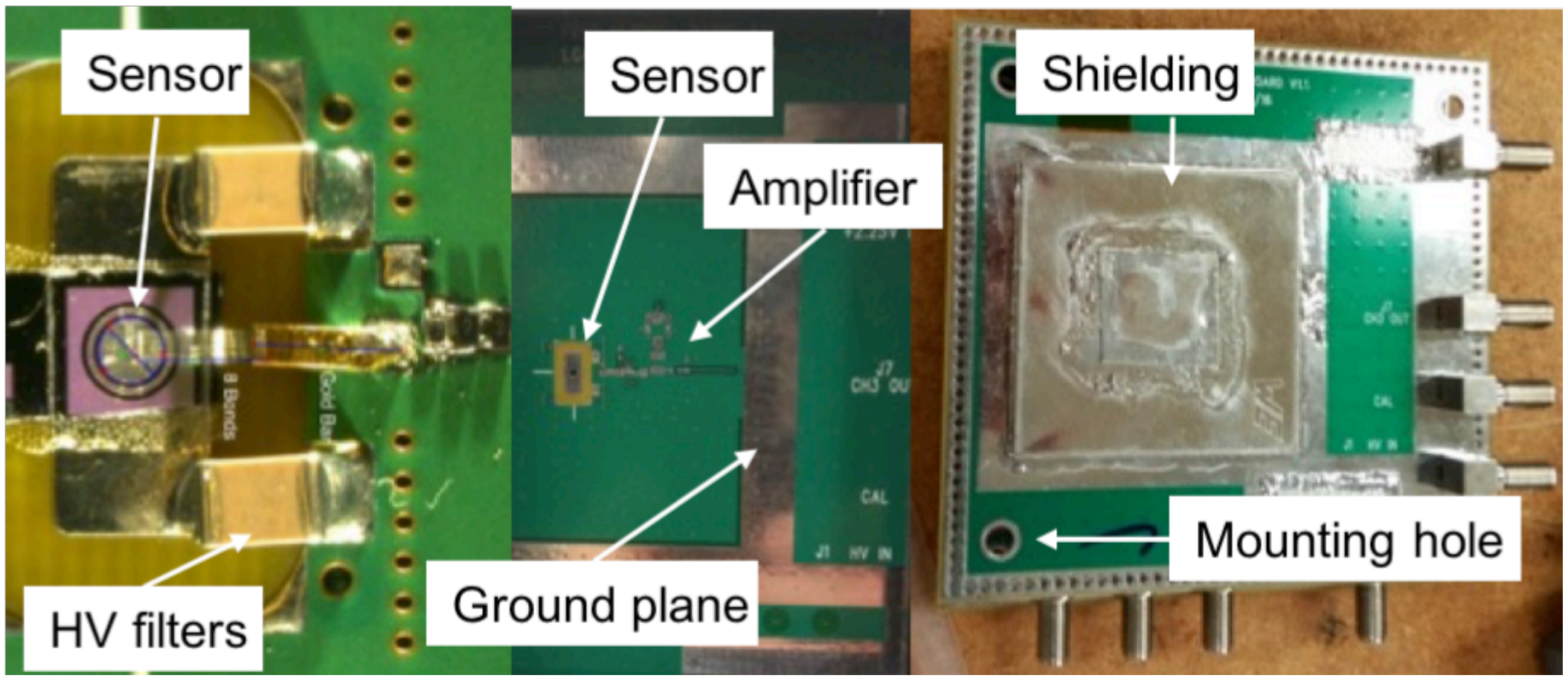

Figure 2 Read-out board: connections to the UFSD (left), board without shielding (centre), board with shielding (right)

## 4 CERENKOV TRIGGER COUNTER

For accurate triggering at beam tests and in the laboratory, a fast Cerenkov trigger counter, based on the concepts of M. Albrow et al. [13], was developed. As shown in Figure 3, it consists of a 10 mm long, 9 $mm^2$ wide quartz bar read out by a 9 $mm^2$ SensL SiPM incorporated in its evaluation board [14]. The SiPM signals were boosted by an amplifier with gain 10 [15].
The UFSD – trigger counter combination was extensively tested in a β-source prior to the beam test, and a biasing voltage range for the SiPM of 28 - 29 V was selected, at the upper limit of the recommended range, 24 to 29 V. The laboratory setup with $^{90}Sr$ β-source allows measuring triple coincidences between the Cherenkov trigger counter and two UFSD detectors: for this set-up, the observed time resolution was limited by the low photon yield of less than 10 due to the low-energy β-particles. No changes in either SiPM noise or thermal photoelectron spectrum were observed between operations in the β-source and the beam. During the beam test, based on the observed signal pulse height and the pulse height spectrum of background photoelectrons, the number of photoelectrons was estimated to be in excess of 50. The trigger threshold was set at 4-5 photoelectrons on a pulse rise time of about 800 ps, which necessitated a time walk correction in the analysis.

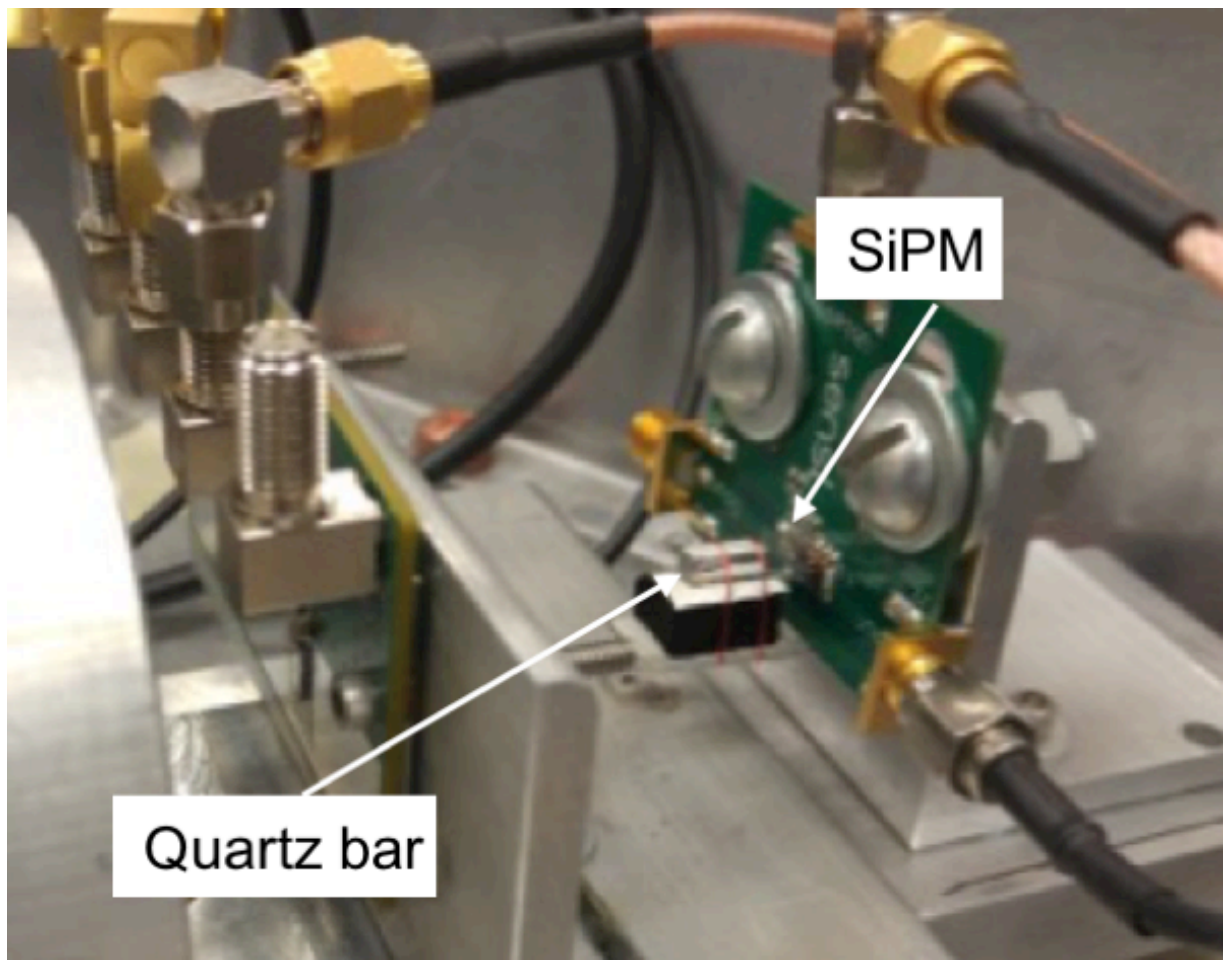

Figure 3 Trigger counter configuration shown in the β-source: the 10mm long quartz bar mounted on the SensL SiPM evaluation

## 5 UFSD Operations at the CERN H8 Test beam

The beam test was conducted in the CERN H8 beam test area with π-mesons with a momentum of 180 GeV/c. The devices under test and the trigger counter were mounted in a frame, Figure 4, having slots for mechanical stability and rough alignment. The SiPM trigger counter was mounted on a G10 plate matching the dimensions of the UFSD readout boards. The fine alignment was accomplished by four precision rods pushed through machined holes at the four corners of the UFSD boards, trigger plate and frame. Limited by the four channels of the digital scope used for the data acquisition, the standard running configuration was three 45 μm UFSD and the trigger counter.

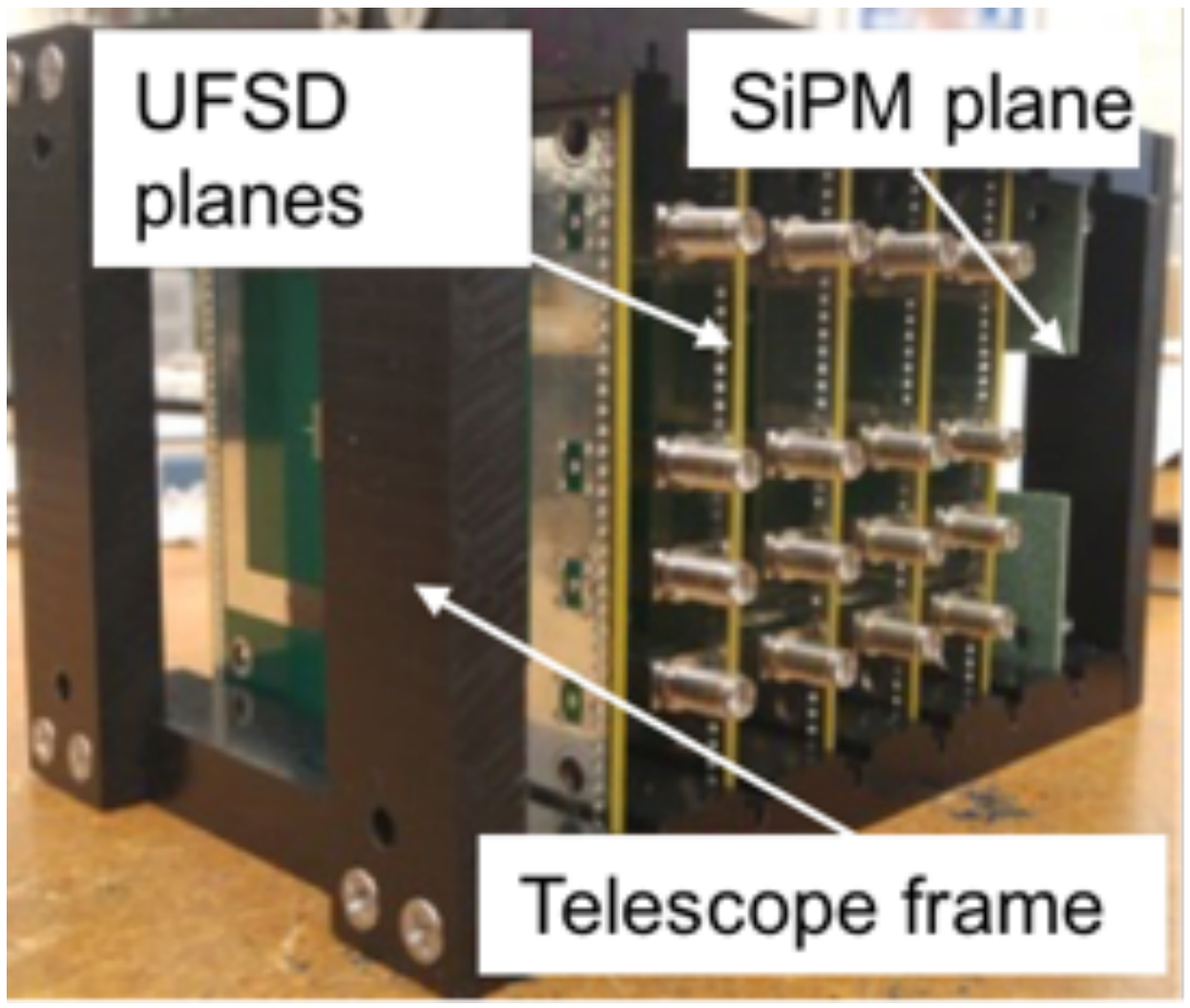

Figure 4 Beam Test frame showing four UFSD boards and the trigger counter. Alignment was assured with four rods (not shown) going through aligning holes in the four corners of the boards and the frame.

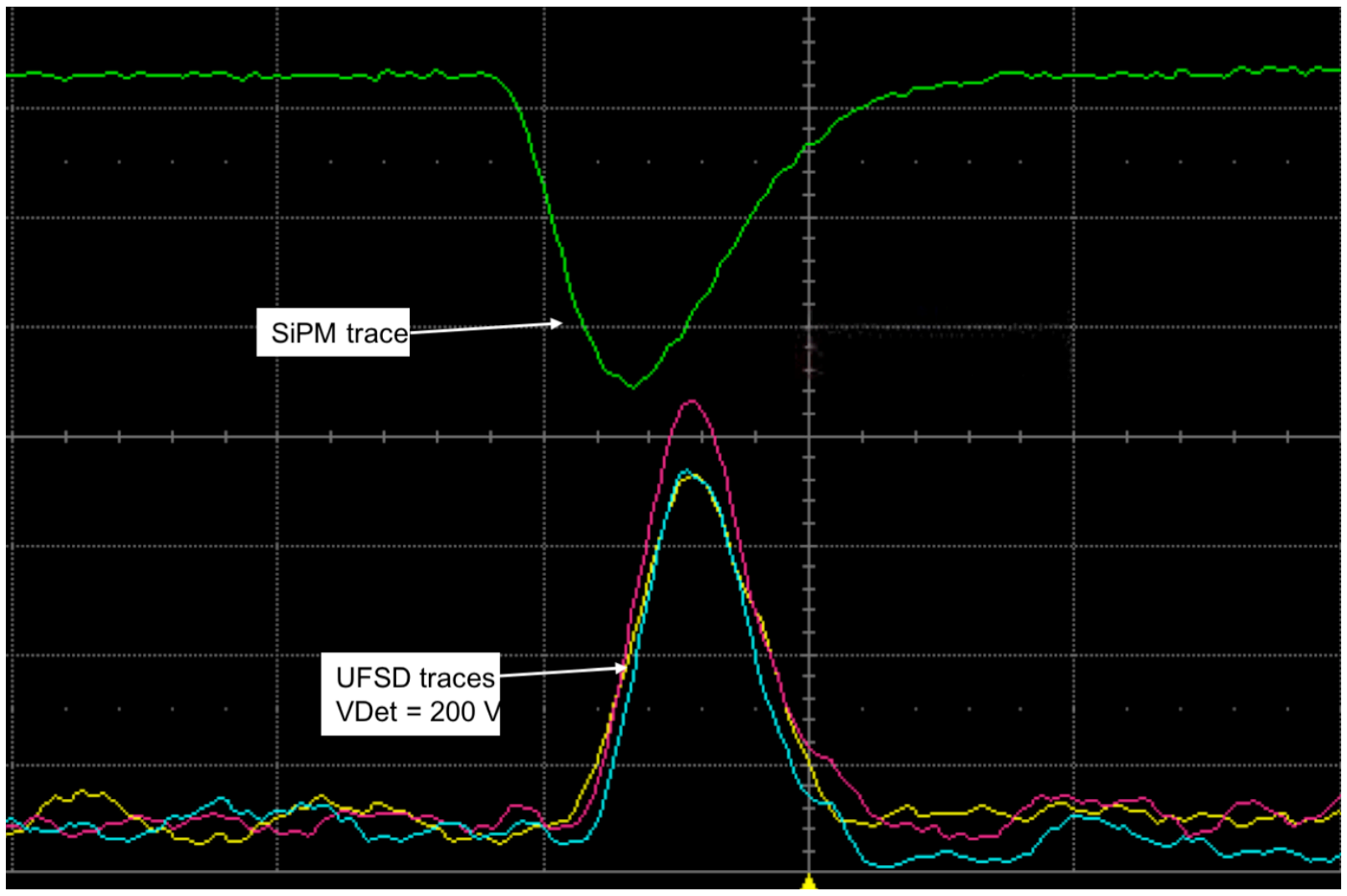

Figure 5 Screenshot of one event, showing the signals of 3 UFSD biased at 200 V and the SiPM at 28 V. Each horizontal division corresponds to 2 ns, while each vertical division is 100 mV.

The data were acquired with a 4 GHz – 8 bit vertical resolution LeCroy WaveRunner Zi digital scope, at a sampling rate of 20 GS/s, therefore with a time discretization of 50 ps. The impact of this time binning on the analysis has been crosschecked acquiring a subset of data with the enhanced data sampling of 40 GS/s and found to be negligible. Figure 5 shows an example of the quality of the data: 3 UFSD traces are visible as positive signals with about 400 ps leading edge time while the trigger counter, with about 600 ps leading edge time, is the negative trace on top. It should be noted that for the UFSD sensors, the observed signal-to-noise ratio S/N and rise time did not change when moving from the β-source in the laboratory to the beam test location. The noise, defined as the r.m.s. of the amplitude distribution obtained combining the first point of each trace, was measured to be in the range 13.5 – 14.0 mV, stable throughout beam test. The traces shown in Figure 5 give an interesting indication about the achievable time resolution: as the signal time jitter can be approximated by (rise time)/(S/N) [16], given the very short rise time (~ 400 ps) even a moderate signal-to-noise ratio of S/N = 20 will yield a time jitter of 20 ps.

Data were collected in spills of 5 sec duration, and an instantaneous trigger rate between 1-20 Hz, depending on the beam conditions. The main objective of the beam test is the evaluation of the time resolution for minimum ionizing particles as a function of the sensor bias voltage. Most of the runs have been acquired with all three UFSD having the same voltage settings. After initial checkout, long data runs with fixed voltage configuration were taken, with typically between 20k and 100k events. Very stable running conditions were observed throughout the week-long test, with the sensors leakage current stable at around 5-10 nA and the SiPM current at 3 – 4 μA. The full dataset comprises of more than half million events.

# 6 Beam Test results

*Analysis procedure*

The digital oscilloscope records the full waveform of each channel in each event, so the complete event information is available to offline analysis. Even though the recorded data would allow an analysis based on multiple sampling techniques [8], the analysis presented here was performed using only the variables available to an hypothetical read-out chip, i.e. the time at threshold and the signal amplitude as in a constant-fraction-discriminator (CFD) chip, or the rising and falling time at a given threshold, as in a Time-over-Threshold (ToT) system. Therefore the results presented in the paper are not the best possible, but they reflect what can be achieved by a dedicated timing read-out system in a large detector.

The event selection is fairly straightforward: the signal amplitude for all 3 UFSD active in the data taking should be above five time the noise level, and not be saturated by either the scope or the read-out chain. To eliminate the contributions from non-gain events, the time of the pulse maximum has to fall into a window of 1 ns. Typical amplitude ranges are shown in Figure 6.

*Gain and Landau distribution*

The initial energy deposition by impinging MIPs follows the Landau distribution, and these initial charges are sub sequentially amplified by the UFSD gain mechanism. The impulse response of the UFSD – amplifier board shown in Figure 2 has been measured in the laboratory using light pulsed from a 1064-nm picolaser, and it was found to be a Gaussian curve with a sigma between 20 and 30 mV, with larger values at higher Vbias. During this study it was also measured that the preamplifier becomes non linear for amplitudes above 800 mV, causing a reduction of the Landau high-energy tail. The non-linearity is of the order of 20% (50%) for amplitudes around 800 mV (1200 mV). Figure 6 shows the beam test measured amplitude distributions at 4 values of Vbias (100 V, 130 V, 200 V, and 230 V) with superposed the result from a Landau-Gauss convolution fit: at Vbias = 100 V and 130 V, the Gaussian contribution to the convolution fit has an amplitude of 22 mV and 28 mV respectively, consistent with the laboratory measurements, while for higher values of Vbias the Gaussian contribution becomes more important as a consequence of the electronics non linearity. At Vbias = 100 V and 130 V the ratio FWHM/MPV of the Landau component is ~ 0.4, somewhat lower than the expected values of ~ 0.6 from e-h pairs creation in thin sensors [11], while for higher Vbias this ratio becomes much smaller due the electronics non-linearity.

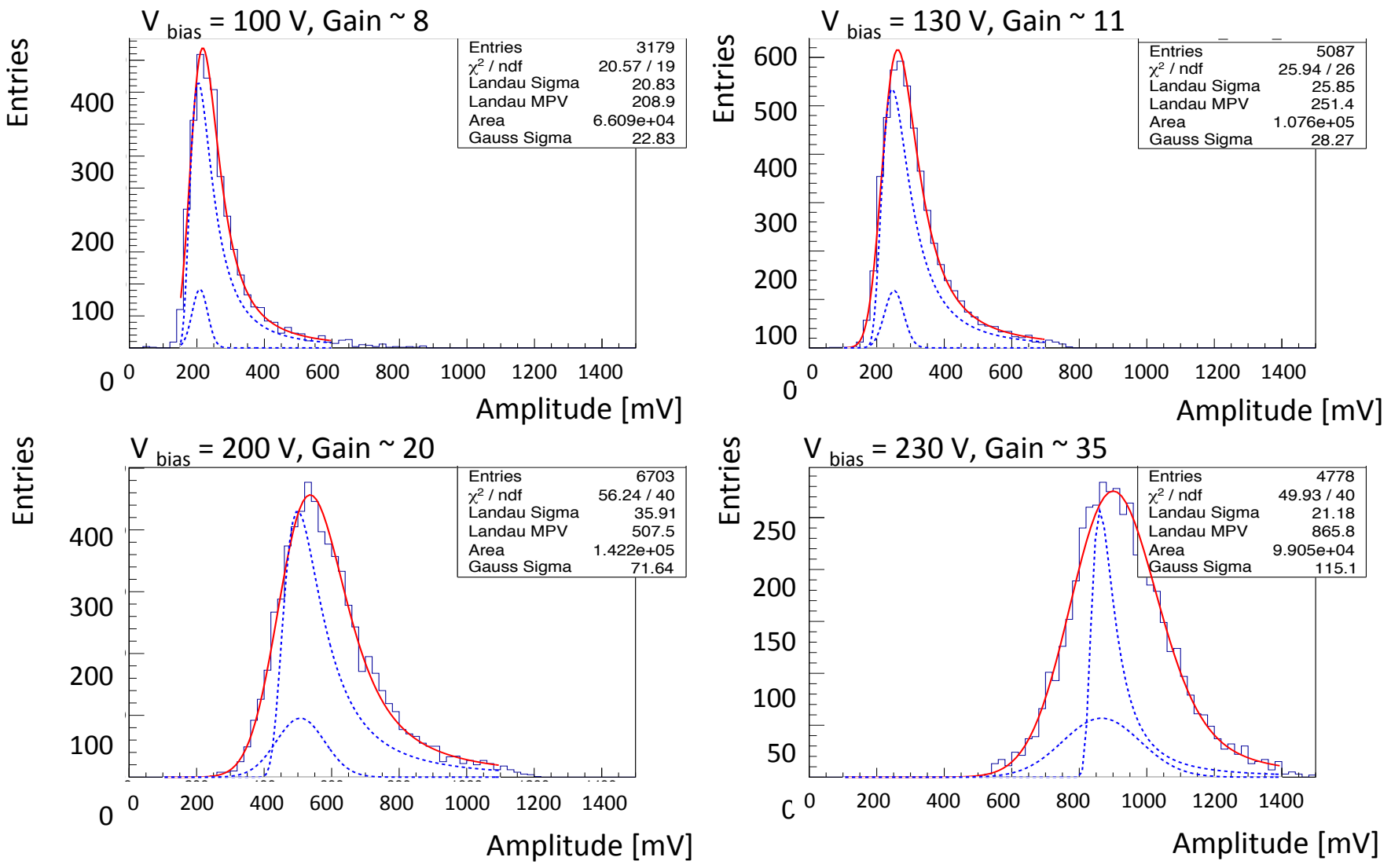

Figure 6 Distribution of the signal amplitude at four values of the detector Vbias: (100V,130V , 200V and 230V), with superposed the curves from a Landau-Gauss convolution fit (solid red curve). On each graph, the normalizations of the Landau and Gauss components are arbitrary as only their product is defined. Likewise, for explanation purpouses only, the gaussian curve is arbitrarly aligned with the Landau MPV value.

*Timing resolution*

The time of passage of a particle in a sensor is evaluated using the constant fraction algorithm (CFD), which defines the time of arrival of a particle as the time at which the signal crosses a certain fraction of the total signal amplitude. The amplitude has been determined with a parabolic fit to the signal 5 top points (spanning 250 ps), while the value of the highest point has been used as a cross check. The time of arrival at a specific fraction value was extracted from the recorded samples by means of a linear interpolation between the point above and below the requested value. By measuring the timing resolution as a function of the CFD fraction, it was concluded that the time resolution is separately minimized for values of CFD between 20% and 30% for each of the UFSD. The same is true for the CFD fraction of the trigger, which was checked separately.

The time resolution of each UFSD and that of the SiPM has been obtained for each bias voltage from an overall fit to the Gaussian sigma's of the 10 distributions of a time difference in one event. There are three time differences between pairs of UFSD ($\Delta(t_{UFSD}-t_{UFSD})$), and 3 time differences between each UFSD and the SiPM ($\Delta(t_{UFSD}-t_{SiPM})$). These differences are based on single UFSD and are called "singlets", In addition one can calculate the three time differences between the average time of two UFSD <2 UFSD> and the SiPM ($\Delta(t_{<2\ UFSD>}-t_{SiPM})$), called "doublets", and finally the time difference between the average of 3 UFSD < 3 UFSD> and the SiPM ($\Delta(t_{<3\ UFSD>}-t_{SiPM})$), the "triplet" . Figure 7 shows the time difference between one of the UFSD and the SiPM at a V bias of 200 V and 230 V. The curve at 200 V shows small non-Gaussian tails (< 1%) while the curve at 230 V has a pronounced low energy tail due to the miscalculation of the CFD point (too low, therefore too early) caused by the saturation of the electronics.

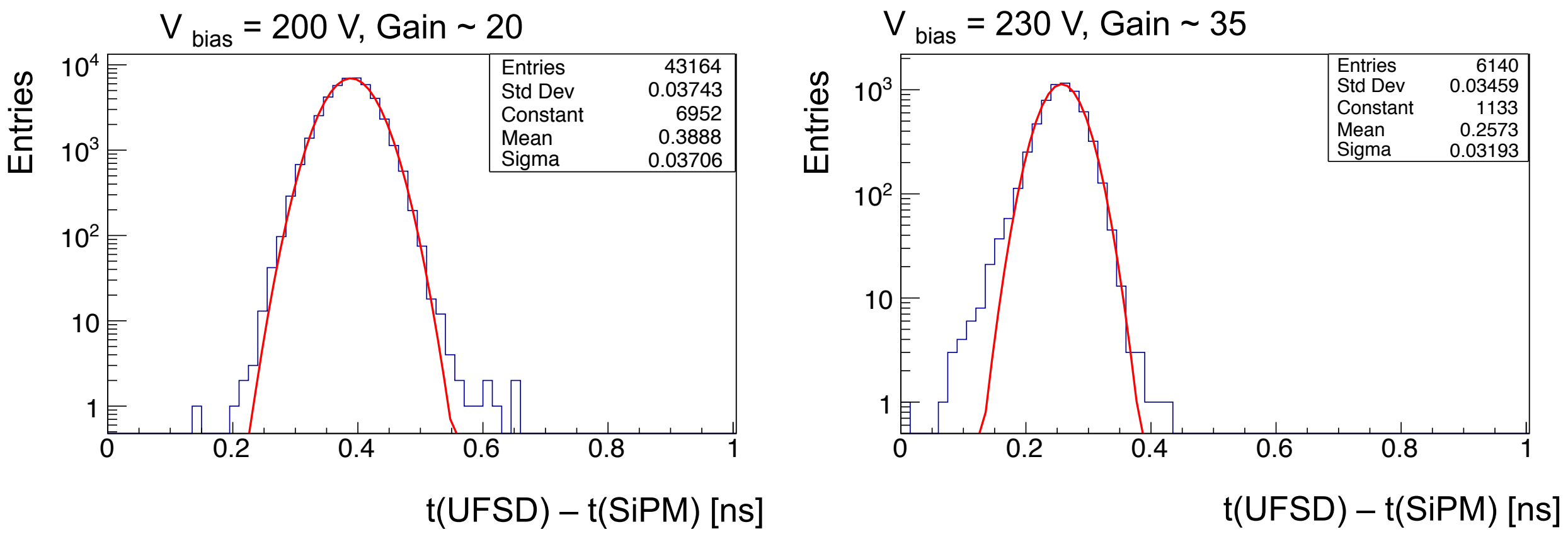

Figure 7 Time difference UFSD-SiPM at 2 different UFSD bias voltages: 200 V with sigma = 37 ps, and 230 V with sigma = 32 ps.

The results for the overall fits are shown in Table 1 and in Figure 8, where all measurements of the timing resolution for singlet, doublets and triplet of UFSD are shown (these values are without SiPM contribution).
The timing resolution of the trigger counter is evaluated to be 13 ps in the run at Vbias = 200 V and 15 ps in the run at Vbias = 230V.

The time resolution of a single UFSD is measured to decrease with increased gain M like $M^{-0.36}$.

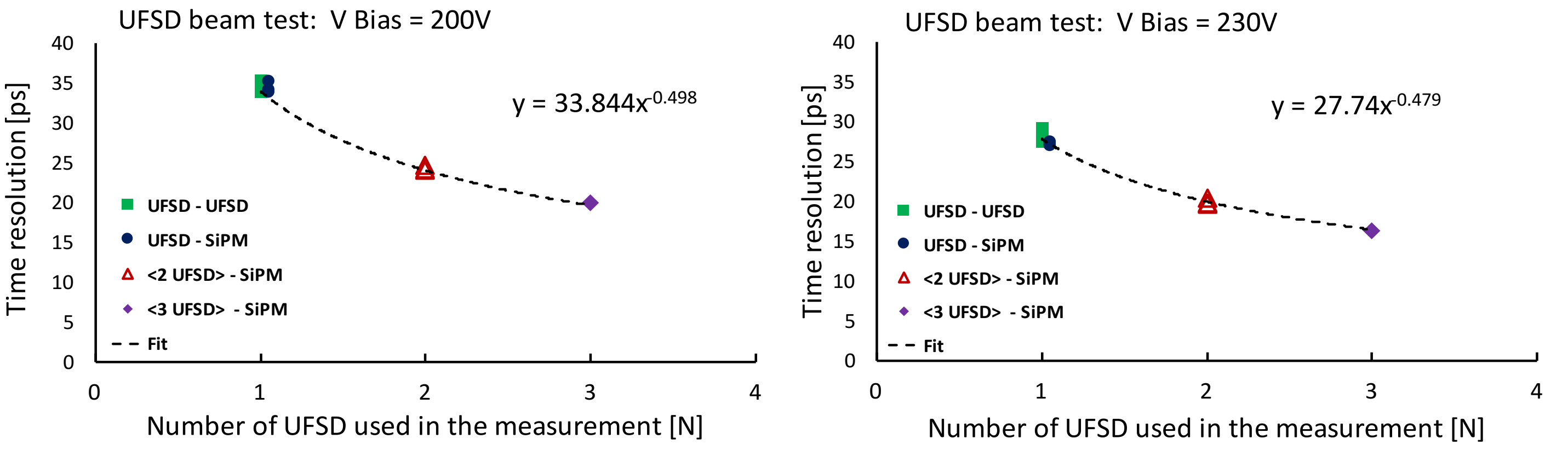

Figure 8 Timing resolution for: single UFSD (N=1) obtained from the UFSD-SiPM and UFSD-UFSD time differences; averaged pairs of UFSD (N=2) and average of 3 UFSD (N = 3) for bias voltages of (left) 200V and (right) 230V.

The scatter of the 6 singlet (N = 1) and 3 doublet (N=2) points in Figure 8 reflects the errors of the fits as well as the gain matching of the UFSD, which appears to be excellent. The scatter is slightly larger for the bias voltage of 230 V since at that voltage the gain is rising rapidly and a difference in gain among the sensors could be expected. Both data set of Figure 8 agree well with the expected σ(N) = $1/\sqrt{N}$ behaviour, demonstrating that the 3 sensors are of equal high quality. The timing resolution of a single UFSD is measured to be 34 ps for 200 V bias and 27 ps for 230 V bias. A system of three UFSD has a measured timing resolution of 20 ps for a bias of 200 V, and 16 ps for a bias of 230 V.

Table 1 Averages of timing resolution for single UFSD (N=1), and for pairs of UFSD (N=2), and the timing resolution for the triplet UFSD (N=3) for two bias voltages. The trigger counter resolution has been subtracted in quadrature to obtain the quoted results.

| **UFSD Timing resolution** | | |
|---|---|---|
| | **Vbias = 200 V** | **Vbias = 230 V** |
| **N = 1** | 34 ps | 27 ps |
| **N = 2** | 24 ps | 20 ps |
| **N = 3** | 20 ps | 16 ps |

Dividing each V bias value in 3 amplitude intervals (250 – 500 mV, 500 – 700 mV, and 700 – 1100 mV for Vbias = 200 V and 250 – 800 mV, 800 – 1000 mV, and 1000 – 1200 mV for Vbias = 230 V), we evaluated that the electronics non linearity worsen the measured time resolution by about 10%. The effect is mostly driven by the incorrect evaluation of the CFD point, since for saturated events the true amplitude is unknown.

## 7 Conclusions

The result of the beam test with pions of momentum 180 GeV/c demonstrates that UFSD with an active thickness of 45 µm and 1.7 $mm^2$ pad size reaches a timing resolution of 34 ps at a bias voltage of 200 V and 27 ps at 230 V. This result is in excellent agreement with what was predicted 3 years ago for thin UFSD by the Weightfield2 simulation program [1]. In the beam test, it was also proved that a telescope comprising 3 planes of UFSD sensors reaches the expected performance improvement given by multiple measurements, i.e. the resolution improves like $1/\sqrt{N}$. For a triplet of UFSD, a timing resolution of 20 ps at a bias voltage of 200 V was achieved, while at 230 V the timing resolution was measured to be 16 ps.

## 8 Acknowledgements

We acknowledge the expert contributions of the SCIPP technical staff. Part of this work has been performed within the framework of the CERN RD50 Collaboration. We thank Michael Albrow and Anatoly Ronzhin for supporting us in the construction of the trigger counters and the TOTEM experiment for the help in the beam test set-up.

The work was supported by the United States Department of Energy, grant DE-FG02-04ER41286. Part of this work has been financed by the Spanish Ministry of Economy and Competitiveness through the Particle Physics National Program (FPA2015-69260-C3-3-R and FPA2014-55295-C3-2-R), by the European Union's Horizon 2020 Research and Innovation funding program, under Grant Agreement no. 654168 (AIDA-2020) and Grant Agreement no. 669529 (ERC UFSD669529), and by the Italian Ministero degli Affari Esteri and INFN Gruppo V.